\documentclass[fleqn,usenatbib,useAMS]{mnras}

\usepackage{graphicx}	% Including figure files
\usepackage{amsmath}	% Advanced maths commands
\usepackage{amssymb}	% Extra maths symbols
\usepackage{multicol}        % Multi-column entries in tables
\usepackage{bm}		% Bold maths symbols, including upright Greek
\usepackage{pdflscape}	% Landscape pages

\usepackage{xcolor}
\usepackage{lipsum}
\usepackage{float}
\usepackage[T1]{fontenc}
\usepackage{ae,aecompl}

\usepackage{newtxtext,newtxmath}
\title[Thin accretion disk luminosity and its image around rotating black holes in perfect fluid dark matter]{{Thin accretion disk luminosity and its image around rotating black holes in perfect fluid dark matter}}

\author[M. Heydari-Fard et al.]{
Malihe Heydari-Fard,$^{1}$\thanks{\href{mailto: heydarifard@qom.ac.ir}{heydarifard@qom.ac.ir}}
Sara Ghassemi Honarvar,$^{2}$\thanks{\href{mailto: s.ghassemihonarvar@mail.sbu.ac.ir}{s.ghassemihonarvar@mail.sbu.ac.ir}}
Mohaddese Heydari-Fard$^{2}$\thanks{\href{mailto: m\_heydarifard@sbu.ac.ir}{m\_heydarifard@sbu.ac.ir}}
\\
$^{1}$Department of Physics, The University of Qom, 3716146611, Qom, Iran\\
$^{2}$Department of Physics, Shahid Beheshti University, Evin 19839, Tehran, Iran
}

\date{Accepted XXX. Received YYY; in original form ZZZ}
\pubyear{2022}

\begin{document}
\label{firstpage}
\pagerange{\pageref{firstpage}--\pageref{lastpage}}
\maketitle

\begin{abstract}
Motivated by the fact that the universe is dominated by dark matter and dark energy, we consider rotating black holes surrounded by perfect fluid dark matter and study the accretion process in thin disk around such black holes. Here, we are interested in how the presence of dark matter affects the properties of the electromagnetic radiation emitted from a thin accretion disk. For this purpose, we use the Novikov-Thorn model and obtain the electromagnetic spectrum of an accretion disk around a rotating black hole in perfect fluid dark matter and compare with the general relativistic case. The results indicate that for small values of dark matter parameter we considered here, the size of the innermost stable circular orbits would decrease and thus the electromagnetic spectrum of the accretion disk increases. Therefore, disks in the presence of perfect fluid dark matter are hotter and more luminous than in general relativity. Finally, we construct thin accretion disk images around these black holes using the numerical ray-tracing technique. We show that the inclination angle has a remarkable effect on the images, while the effect of dark matter parameter is small.
\end{abstract}


%%%%%%%%%%%%%%%%%%%%%%%%%%%%%%%%%%%%%%%%%%%%%%%%%%

%%%%%%%%%%%%%%%%% BODY OF PAPER %%%%%%%%%%%%%%%%%%

\section{Introduction}
Black holes are fascinating extreme objects predicted by the theory of general relativity (GR). They are known as the strongest sources of gravitational field in the universe and typically expected to have very high spin and magnetic intensity. Due to such features, black holes are an ideal laboratory for studying both matter and gravity in astrophysical experiments. In recent years, some observational evidence have confirmed the existence of black holes. The first was the discovery of gravitational waves resulting from a binary black hole merger by LIGO and Virgo collaboration \citep{LIGOScientific:2016aoc}. Another extraordinary milestone is the first image of M87* black hole shadow \citep{EventHorizonTelescope:2019dse,EventHorizonTelescope:2019ths} and recently released image of Sgr A* \citep{EventHorizonTelescope:2022wkp} by the Event Horizon Telescope using Very Long Baseline Interferometry. In addition, the observed electromagnetic spectrum of accretion disks can provide confirmation of the existence of black holes \citep{frank_king_raine_2002,Yuan:2014gma,2018arXiv181007041N}. These recent achievements provide an enlightening response to the understanding of the theory of GR and the nature of accretion disks around supermassive compact objects in the strong gravity regime near the event horizon and may be considered as a way for testing modified theories of gravity.

It is generally believed that compact objects such as black holes gain their mass from the accretion process. Also, to preserve the high accretion rate around such compact objects, the existence of accretion disk is an essential ingredient. An accretion disk is formed by diffusing material which slowly spiral into central compact object and releases gravitational energy as radiation. Images of accretion disks around black holes have been one of the most interesting research areas in observational astronomy since 1970s. The standard model of geometrically thin and optically thick accretion disks was initially proposed by Shakura and Sunyaev in 1973 \citep{shakura1973black} and later generalised by Page, Novikov and Thorne \citep{novikov1973black,Page:1974he} to include general relativity which successfully explains the observed spectral features of astrophysical black holes. Also, for a review on theoretical models of black hole accretion disks, see \citep{Abramowicz:2011xu}. The signatures appearing in the energy flux and emission spectra of disks provide us not only with information about central massive object, but also provide a test for modified theories of gravity. So, the study of accretion disks around black holes may be considered as a tool to explore possible deviations of alternative theories from GR. The properties of thin disks in various background space-times have been extensively studied
\citep{Kazempour:2022asl,Liu:2020vkh,Liu:2021yev,2022arXiv220803723H,Boshkayev:2021chc,Gyulchev:2021dvt,Heydari-Fard:2010agr,Heydari-Fard:2020iiu,Heydari-Fard:2020ugv,Heydari-Fard:2021ljh, Zhang:2021hit,Karimov:2018whx,Karimov:2019qfw,Karimov:2020fuj,Jawad:2019oxl,Perez:2012bx,Perez:2017spz,Chen:2011wb,Chen:2011rx,Harko:2009rp,Harko:2009kj,Guzman:2005bs,1986AN....307..183P}

As we know, according to the standard model of cosmology the matter-energy content of our universe consist of $68\%$ dark energy and $27\%$ dark matter, while the contribution of baryonic matter is only about $5\%$ of the total matter-energy of the universe \citep{jarosik2011seven}. Although there has been no direct detection of dark matter, there are some observational evidence confirming its existence including the asymptotic flatness of rotation curves of giant spiral and elliptical galaxies \citep{Rubin:1980zd}, the observations related to the dynamics of galaxy clusters \citep{Zwicky:1933gu} and also the measurements of cosmic microwave background (CMB) anisotropies \citep{Planck:2013lkt}. There are different models of dark matter used to explain large and small structure of the universe \citep{Navarro:1995iw,Tulin:2017ara,Kiselev:2004vy}. In most of these models the dark matter is assumed as a halo around black holes \citep{Jusufi:2019nrn,Konoplya:2019sns}. Moreover, the quintessential dark matter solutions was initially proposed by Kiselev \citep{2003gr.qc.....3031K,Kiselev:2002dx}. Also, an interesting model was developed by Rahaman et al. in which the black hole is surrounded by perfect fluid dark matter (PFDM) with energy-momentum tensor given by $T_{\mu\nu}=(\rho+p)u_{\mu}u_{\nu}+pg_{\mu\nu}$ \citep{Rahaman:2010xs}. Later, this PFDM model was revisited in the light of the observed data of Milky Way galaxy \citep{Potapov:2016obe} where the authors obtained a range for the state parameter of PFDM that indicates a dust-like cold dark matter halo also supported by CMB constraints \citep{mueller2005cosmological,kumar2014observational}. On the other hand, a model of PFDM around black hole in the background of spatially inhomogeneous phantom field has been proposed in \citep{Li:2012zx}. The Reissner-Nordstrom-anti-de Sitter black hole surround by PFDM has been obtained in \citep{xu2019perfect}. Also, the extension of black hole solutions in the presence of PFDM to the case of rotating black hole solutions is done in \citep{Xu:2017bpz} and the behavior of rotational velocity with respect to dark matter parameter and black hole spin have also been investigated. The shadow of rotating black holes in PFDM with and without cosmological constant was studied in \citep{Haroon:2018ryd,Hou:2018avu}. Moreover, the analysis of the shadow cast by rotating charged black holes in the presence of PFDM has been investigated in \citep{Atamurotov:2021hck,Das:2021otl}. The study of circular geodesics in the space-time of rotating charged black holes in PFDM have also been done in \citep{Das:2020yxw}.

According to the above discussions, it is natural to explore the effects of dark matter on accretion disk luminosity. Such studies, including the luminosity of accretion disks in the space-time of a static black hole in a dark matter envelope and also of  black holes surrounded by dark matter with isotropic/anisotropic pressure, have been investigated in \citep{Boshkayev:2020kle,Boshkayev:2022vlv,Kurmanov:2021uqv}. Now, in the present work we are going to consider rotating black holes surrounded by PFDM \citep{Xu:2017bpz} and study the influence of dark matter on the electromagnetic properties of thin disks such as the radiant energy flux, temperature distribution, observed luminosity and also radiative efficiency. The results are compared to those of rotating black holes in the absence of dark matter in GR.

The paper is organized as follows. In section~\ref{2-PFDM BH}, we briefly introduce the rotating black holes in PFDM and explain some of their properties. The geodesic motion of test particles in the space-time of such black holes is studied in section~\ref{3-geodesic}. Employing the Novikov-Thorne model, we investigate the electromagnetic properties of thin accretion disks around rotating black holes surrounded by PFDM in section~\ref{4-disk}. In section~\ref{5-image} we plot the ray-traced redshifted image and intensity and polarization profile of an accretion disk around rotating black holes in PFDM. Finally, we summarized our results in section~\ref{6-conclusion}.

\section{Rotating black holes in perfect fluid dark matter}
\label{2-PFDM BH}
\begin{figure}
\centering
\includegraphics[width=\columnwidth]{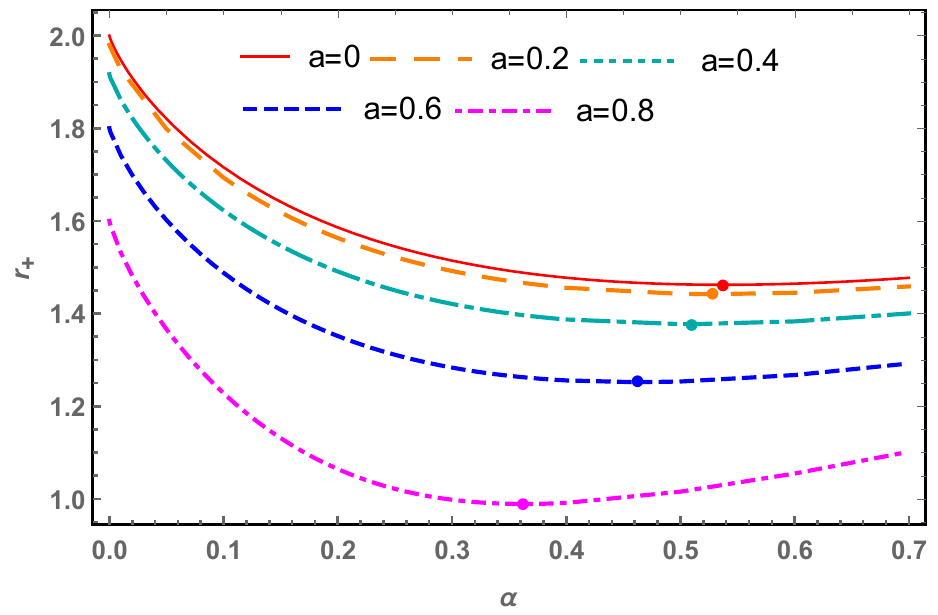}
\caption{Behavior of the  event horizon radius as a function of PFDM parameter $\alpha$. The location of critical value $\alpha_{c}$ is indicated in each case.}
\label{event-horizon}
\end{figure}

We consider that the central object (black hole) is surrounded by dark matter so that the dark matter field is minimally coupled to gravity \citep{2003gr.qc.....3031K,Kiselev:2004vy,Li:2012zx}
\begin{equation}
{\cal S}=\int d^4x \sqrt{-g}\left(\frac{R}{16\pi}+{\cal L}_{\rm DM}\right),
\label{n1}
\end{equation}
where $g$ is the determinant of the metric tensor, $R$ is the Ricci scalar, and ${\cal L}_{\rm DM}$ gives the Lagrangian density for any dark matter field. Note that there is no interaction between dark matter and ordinary matter. Varying the action gives the field equations as
\begin{equation}
R_{\mu\nu}-\frac{1}{2}g_{\mu\nu}R=8\pi T^{\rm DM}_{\mu\nu},
\label{n2}
\end{equation}
where $T^{\rm DM}_{\mu\nu}$ corresponds to the energy-momentum tensor of dark matter. In this work, we assume that the surrounding dark matter is of PFDM kind with the energy-momentum tensor given by $T^{\mu}_{\nu}=\rm diag(-\rho,p,p,p)$ and $\rho$ and $p$ being the density and pressure, respectively. Assuming that the equation of state for PFDM to be $p=(\delta-1)\rho$, where $\delta$ is a constant, the static spherically symmetric black hole solution is obtained as follows \citep{2003gr.qc.....3031K,Li:2012zx}
\begin{eqnarray}
ds^2&=&-\left(1-\frac{2M}{r}+\alpha r\ln(\frac{r}{|\alpha|})\right)dt^2+\frac{dr^2}{\left(1-\frac{2M}{r}+\alpha r\ln(\frac{r}{|\alpha|})\right)}\nonumber\\
&+&r^2(d\theta^2+\sin^2\theta d\phi^2),
\label{n3}
\end{eqnarray}
where $M$ is the black hole mass and $\alpha$ is the dark matter parameter describing the intensity of PFDM. It has been shown that the logarithmic term  explains the asymptotically flat rotation velocity in spiral galaxies. Also, in the limit of $\alpha\rightarrow0$ the metric corresponds to the Schwarzschild solution. The line element of a rotating black hole surrounded by PFDM is also given by \citep{Xu:2017bpz}
\begin{eqnarray}
ds^2&=&-\frac{1}{\rho^2}\left(\Delta-a^2\sin^2\theta\right)dt^2+\frac{\rho^2}{\Delta}{dr^2}+\rho^2 d\theta^2\nonumber\\
&-&\frac{2a\sin^2\theta}{\rho^2}\left[2Mr-\alpha r\ln(\frac{r}{|\alpha|})\right]dt d\phi+\sin^2\theta\nonumber\\
&\times&\left[r^2+a^2+\frac{a^2\sin^2\theta}{\rho^2}\left(2Mr-\alpha r\ln(\frac{r}{|\alpha|})\right)\right] d\phi^2,
\label{1}
\end{eqnarray}
with
\begin{equation}
\Delta=r^2+a^2-2Mr+\alpha r\ln(\frac{r}{|\alpha|}), \quad \rho^2=r^2+a^2\cos^2\theta.
\label{2}
\end{equation}
In the absence of PFDM, namely in the limit of $\alpha\rightarrow0$, the above metric reduces to the Kerr metric and for $a=0$ it represents the line element of a static black hole surrounded by PFDM in equation \ref{n3}. The black hole horizons can be obtained from equation $\Delta=0$. The presence of PFDM does not change the number of horizons and only modifies the location of them. There are two horizons corresponding to inner horizon $r_{-}$, and outer event horizon $r_{+}$ \citep{Das:2020yxw}. The behavior of the event horizon as a function of dark matter parameter for different values of spin parameter is plotted in Fig.~\ref{event-horizon}. Note that in the present work we consider $\alpha$ to take positive values which gives the positive energy density of dark matter, while in \citep{Xu:2017bpz} both positive and negative cases have been discussed. As the figure shows, the event horizon radius decreases with the increase of spin parameter so that for a given value of $\alpha$, the static PFDM black hole with $a=0$ has the largest values of the event horizon. Moreover, we see that for a fixed value of $a$ the event horizon radius initially decreases with increasing  dark matter parameter and increases around a critical value of $\alpha$. This increase in the black hole size may be attributed to the fact that after a critical value, the dark matter contributes to the effective mass of the black hole. Therefore, in what follows we consider a more realistic case in which for values of $\alpha$ less than the critical value, $\alpha_{c}$, the black hole mass overcomes dark matter distribution.

\section{Geodesic equations}
\label{3-geodesic}
The general form of the metric for a stationary and axisymmetric space-time is given by
\begin{equation}
ds^2=g_{tt}dt^2+2g_{t\phi}dtd\phi+g_{rr}dr^2+g_{\theta\theta}d\theta^2+g_{\phi\phi}d\phi^2.
\label{4}
\end{equation}
Since we assume that the metric coefficients are independent of $t$ and $\phi$ coordinates, two constants of motion corresponding to the energy and angular momentum per unit rest-mass, $\tilde{E}$ and $\tilde{L}$, can be obtained from Euler-Lagrange equations
\begin{equation}
g_{tt}\dot{t}+g_{t\phi}\dot{\phi}=-\tilde{E},
\label{5}
\end{equation}
\begin{equation}
g_{t\phi}\dot{t}+g_{\phi\phi}\dot{\phi}=\tilde{L}.
\label{6}
\end{equation}
Now, taking the condition $2{\cal L}=-1$ and using above equations we find
\begin{equation}
g_{rr} \dot{r}^2+g_{\theta\theta}\dot{\theta}^2=V_{\rm eff}(r,\theta),
\label{9}
\end{equation}
where the effective potential is given by
\begin{equation}
V_{\rm eff}(r,\theta)=-1+\frac{\tilde{E}^2g_{\phi\phi}+2\tilde{E}\tilde{L}g_{t\phi}+\tilde{L}^2g_{tt}}{g_{t\phi}^2-g_{tt}g_{\phi\phi}}.
\label{10}
\end{equation}
For circular orbits in the equatorial plane, $\theta=\frac{\pi}{2}$, with conditions $\dot{r}=\dot{\theta}=\ddot{r}=0$ the radial component of the geodesic equation leads to the following relation for the angular velocity $\Omega=\dot{\phi}/\dot{t}$ \citep{Bambi:2017khi}
\begin{equation}
\Omega_{\pm}=\frac{-g_{t\phi,r}\pm\sqrt{(g_{t\phi,r})^2-g_{tt,r}g_{\phi\phi,r}}}{g_{\phi\phi,r}},
\label{11}
\end{equation}
where the upper and lower signs denote co-rotating and counter-rotating orbits, respectively.
Then, using $g_{\mu\nu}\dot{x^{\mu}}\dot{x^{\nu}}=-1$ and equations (\ref{5}) and (\ref{6}) one can find the specific energy $\tilde{E}$ and the specific angular momentum $\tilde{L}$ of test particles moving in circular orbits as
\begin{equation}
{\tilde{E}}=-\frac{g_{tt}+g_{t\phi}\Omega}{\sqrt{-g_{tt}-2g_{t\phi}\Omega-g_{\phi\phi}\Omega^2}},
\label{12}
\end{equation}
\begin{equation}
{\tilde{L}}=\frac{g_{t\phi}+g_{\phi\phi}\Omega}{\sqrt{-g_{tt}-2g_{t\phi}\Omega-g_{\phi\phi}\Omega^2}}.
\label{13}
\end{equation}
\begin{figure}
\includegraphics[width=\columnwidth]{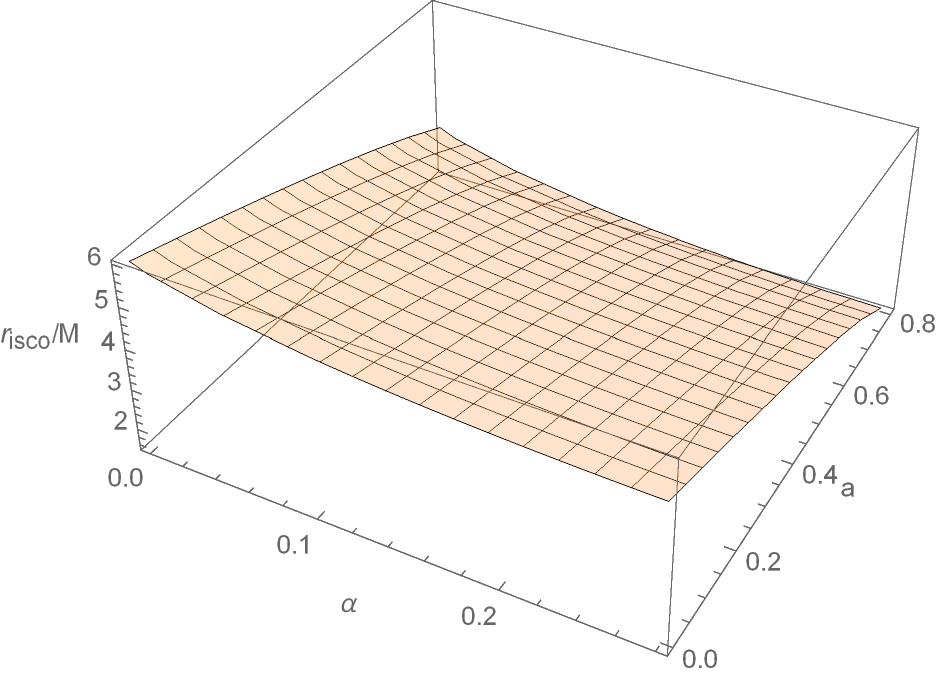}
\caption{Behavior of the ISCO radius as a function of PFDM parameter $\alpha$ and the spin parameter $a$.}
\label{isco-radius}
\end{figure}
Now, using equations (\ref{11})-(\ref{13}) we find the angular velocity, the specific energy and the specific angular momentum for particles moving in circular orbits around a rotating black hole in PFDM as follows
\begin{equation}
\Omega=\frac{1}{a+\sqrt{2}r/h(r)},
\label{14}
\end{equation}
\begin{equation}
{\tilde{E}}=\frac{\sqrt{2}(r-2M+\alpha\ln(\frac{r}{\alpha}))+ah(r)}{\sqrt{r[-6M+2r-\alpha(1-3\ln(\frac{r}{\alpha}))+2\sqrt{2}ah(r)]}},
\label{15}
\end{equation}
\begin{equation}
{\tilde{L}}=\frac{-\sqrt{2}a(2M-\alpha\ln(\frac{r}{\alpha}))+(r^2+a^2)h(r)}{\sqrt{r[-6M+2r-\alpha(1-3\ln(\frac{r}{\alpha}))+2\sqrt{2}ah(r)]}},
\label{16}
\end{equation}
with
\begin{equation}
h(r)=\sqrt{\frac{2M+\alpha(1-\ln(\frac{r}{\alpha}))}{r}}.
\label{17}
\end{equation}
The effective potential  is also given by
\begin{eqnarray}
V_{\rm eff}&=&-1+\frac{1}{r\left[a^2+r(r-2M)+r\alpha\ln(\frac{r}{\alpha})\right]}\times\left(2M(\tilde{L}-a\tilde{E})^2\right.\nonumber\\
&+&\left.r(a\tilde{E}-\tilde{L})(a\tilde{E}+\tilde{L})+\tilde{E}^2r^3-(\tilde{L}-a\tilde{E})^2\alpha\ln(\frac{r}{\alpha})\right),
\label{18}
\end{eqnarray}
where in the absence of PFDM, $\alpha=0$, coincides with that of Kerr black hole. Moreover, the innermost stable circular orbit, $r_{\rm isco}$, around the central compact object can be determined from the condition $V_{\rm eff,rr}=0$, which leads to
\begin{eqnarray}
&&2r(6M^2-Mr+4M\alpha+\alpha^2)-4\sqrt{2}ar(2M+\alpha)h(r)\nonumber\\
&+&3r\alpha^2\ln(\frac{r}{\alpha})^2+a^2(6M+4\alpha)+\alpha\ln(\frac{r}{\alpha})\nonumber\\
&\times&\left(-3a^2+r(r-12M-4\alpha)+4\sqrt{2}arh(r)\right)=0,
\label{20}
\end{eqnarray}
that does not have an analytical solution. We have therefore numerically obtained the ISCO radius for different values of $a$ and $\alpha$ parameters. It is clear that in the case of $a=\alpha=0$ the above equation reduces to the Schwarzschild black hole with $r_{\rm isco}=6M$.
In Fig.~\ref{isco-radius}, we have shown the effect of both the spin parameter and dark matter parameter on the radius of innermost stable circular orbit. We see that by increasing the spin and dark matter parameters the ISCO radius decreases, however the rate of such a decrease is larger for the spin parameter.

\section{Thin accretion disk properties around rotating black holes in perfect fluid dark matter}
\label{4-disk}
In this section, we apply the steady-state Novikov-Thorne model to investigate the properties of thin accretion disks in the presence of PFDM. The model gives a geometric description of thin accretion disk for which its vertical size is negligible compared to its horizontal size, $h\ll r$. The accretion disk is in the equatorial plane of the central massive object and its inner edge is at the ISCO radius. The disk's particles have Keplerian motion between $r_{\rm isco}$ and outer edge $r_{\rm out}$ and the mass accretion rate, $\dot{M}_0$, is assumed to be constant in time \citep{Bambi:2017khi}.
\begin{figure*}
\begin{multicols}{1}
\includegraphics[width=\columnwidth]{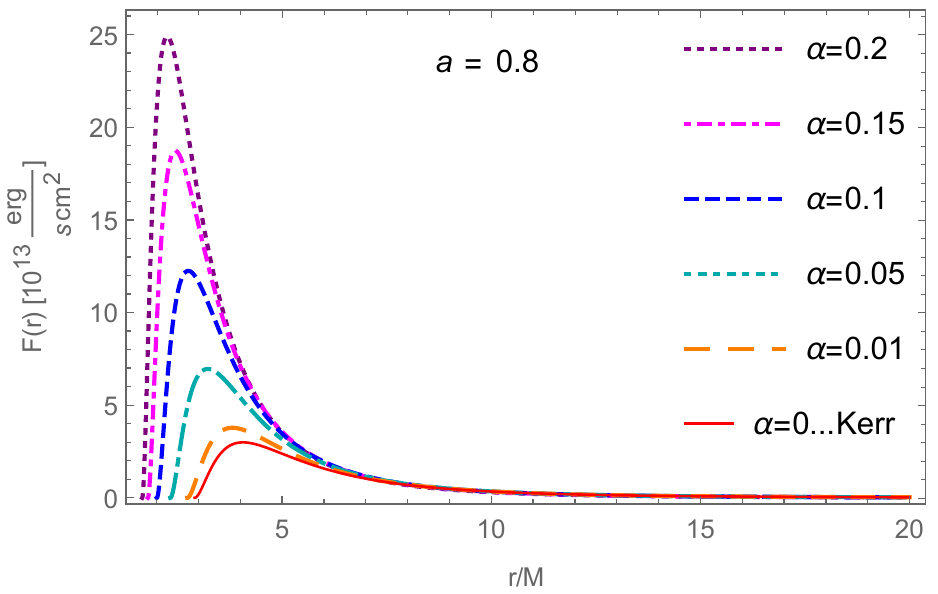}\par
\includegraphics[width=\columnwidth]{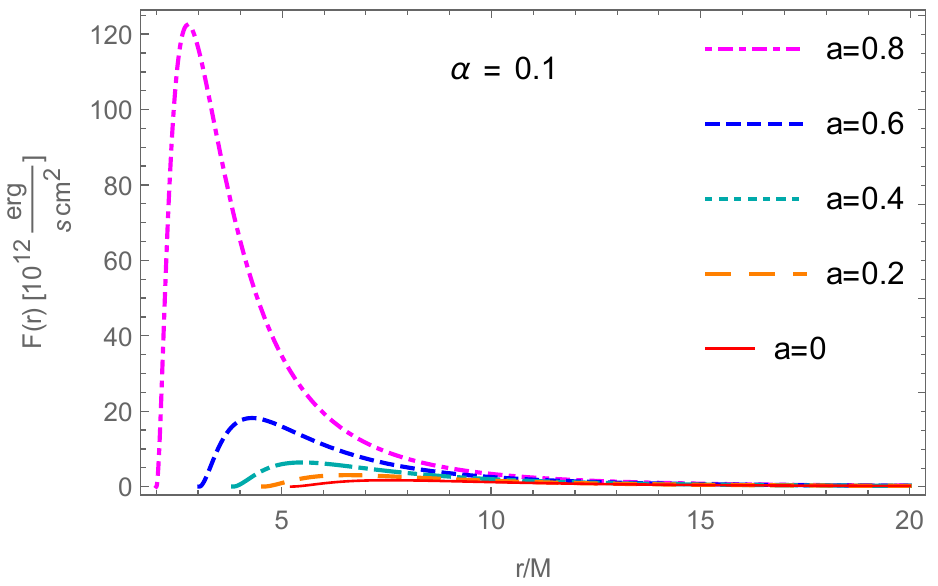}\par
\end{multicols}
\caption{The energy flux $F(r)$ of an accretion disk around the rotating black hole in presence of PFDM with mass accretion rate
$\dot{M}=2\times10^{-6}M_{\odot}yr^{-1}$ for different values of the parameter $\alpha$ in the spin parameter $a=0.8$ (left) and for different values of spin parameter with $\alpha=0.1$ (right).}
\label{flux}
\end{figure*}

\begin{figure*}
\begin{multicols}{1}
\includegraphics[width=\columnwidth]{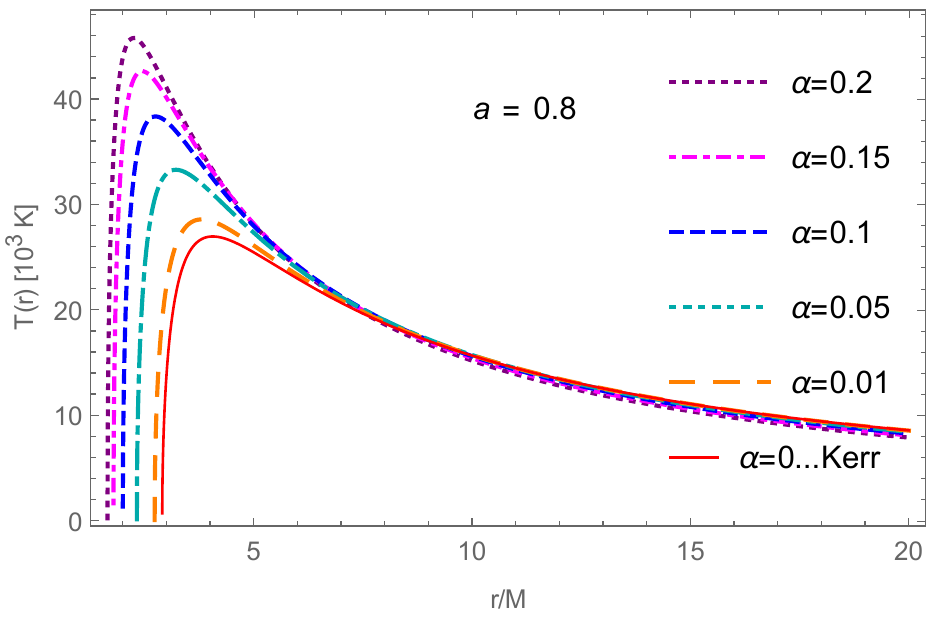}\par
\includegraphics[width=\columnwidth]{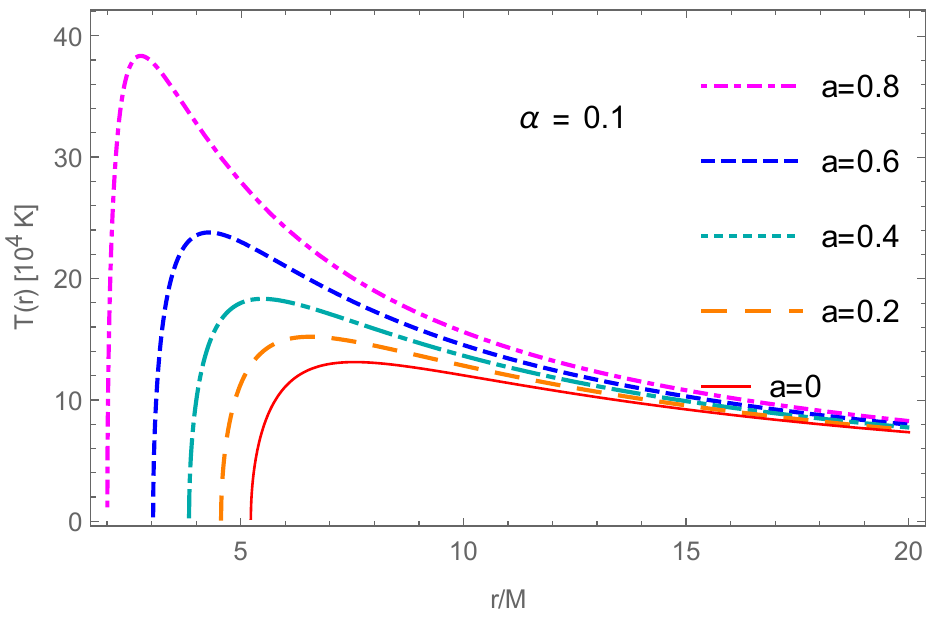}\par
\end{multicols}
\caption{The temperature $T(r)$ of an accretion disk around the rotating black hole in presence of PFDM with mass accretion rate $\dot{M}=2\times10^{-6}M_{\odot}yr^{-1}$ for different values of the parameter $\alpha$ in the spin parameter $a=0.8$ (left) and for different values of the spin parameter with $\alpha=0.1$ (right).}
\label{temperature}
\end{figure*}

From the conservation laws of energy $(\nabla_\mu T^{t\mu}=0),$ angular momentum $(\nabla_\mu T^{\phi\mu}=0),$ and rest mass $(\nabla_\mu(\rho u^\mu)=0)$ of disk particles, one can obtain the radiant energy flux of accretion disk according to \citep{novikov1973black,Page:1974he}
\begin{equation}
F(r)=-\frac{\dot{M}\Omega_{,r}}{4\pi\sqrt{-g}(\tilde{E}-\Omega \tilde{L})^2}\int_{r_{\rm isco}}^r (\tilde{E}-\Omega\tilde{L})\tilde{L}_{,r}dr.
\end{equation}
Since the accretion disk is assumed to be in local thermal equilibrium the disk emission is blackbody radiation for which the Stefan-Boltzmann law is valid
\begin{equation}
F(r)=\sigma_{\rm SB}T^4(r),
\end{equation}
where $\sigma_{\rm SB}$ is the Stefan-Boltzmann constant. Also, the observed luminosity of thin disk has a red-shifted black body spectrum \citep{Torres:2002td}
\begin{equation}
L(\nu)=8\pi h\cos\gamma\int_{r_{\rm in}}^{r_{\rm out}}\int_0^{2\pi}\frac{\nu_e^3 rdrd\phi}{e^{\frac{h\nu_e}{k_B T}}-1},
\end{equation}
where $h$ is the Planck constant, $k_B$ is the Boltzmann constant, and $\gamma$ is the disk inclination angle which we will set to zero. $\nu_e$ is the emitted frequency and is given by $\nu_e=\nu(1+z)$ where $z$ is the redshift factor and can be calculated by
\begin{equation}
1+z=\frac{1+\Omega r\sin\phi\sin\gamma}{\sqrt{-g_{tt}-2g_{t\phi}\Omega- g_{\phi\phi}\Omega^2}}.
\end{equation}

Now, one is ready to study thin accretion disk properties around rotating black holes in PFDM and compare the results with that of Kerr black hole in GR. The behavior of energy flux of an accretion disk around a rotating PFDM black hole for different values of parameter $\alpha$ is plotted in the left panel of Fig.~\ref{flux}. It can be seen that with increasing $\alpha$, the energy flux of the accretion disk also increases. Moreover, the peaks of the energy flux diagram tend towards smaller radii with  increasing $\alpha$, which means that most of the energy flux is emitted from the inner part of the disk. To examine the effect of the spin parameter, we have plotted the energy flux when $\alpha=0.1$ for different values of the spin parameter $a$, in the right panel of the figure. It is clear, with increasing spin parameter, the energy flux also increases, so that the disk around a static PFDM black hole has the smallest energy flux. The disk temperature profile is shown in Fig.~\ref{temperature}, where the same features can be seen.

\begin{figure*}
\begin{multicols}{1}
\includegraphics[width=\columnwidth]{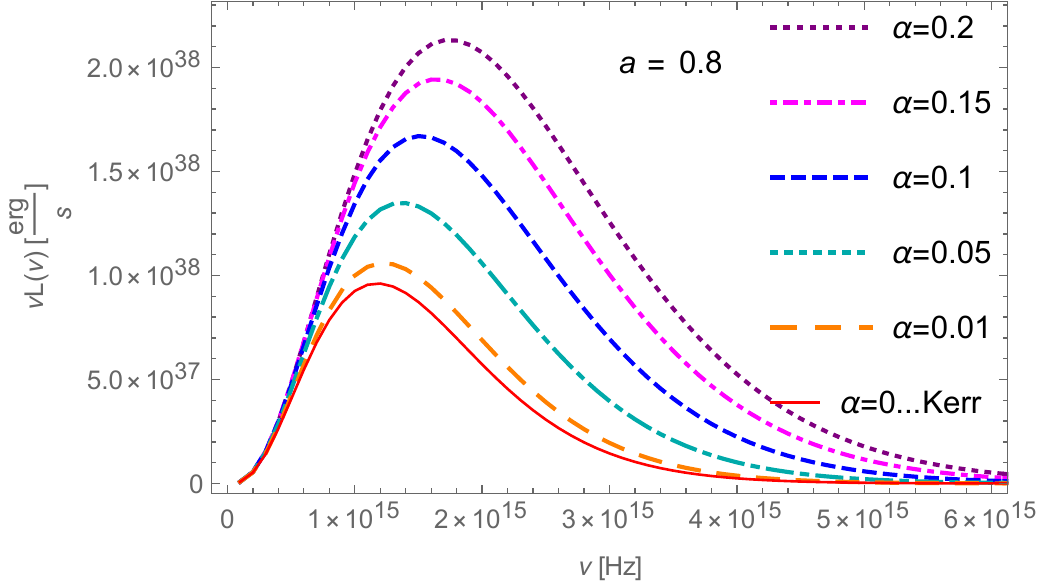}\par
\includegraphics[width=\columnwidth]{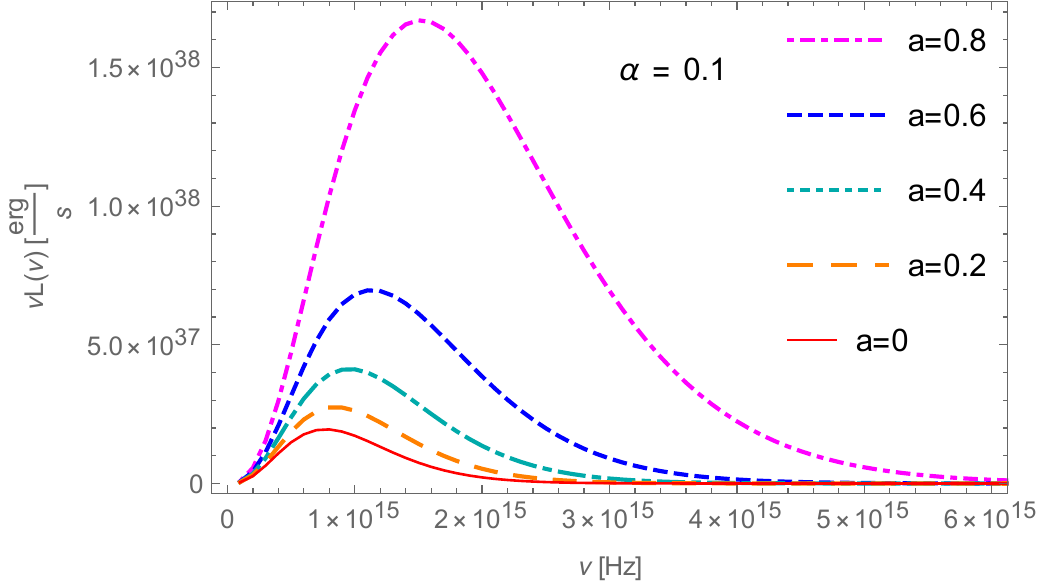}\par
\end{multicols}
\caption{The emission spectrum $\nu L(\nu)$ of an accretion disk around the rotating black hole in the presence of PFDM with mass accretion rate $\dot{M}=2\times10^{-6}M_{\odot}yr^{-1}$ as a function of frequency $\nu$ for different values of the parameter $\alpha$ in the spin parameter $a=0.8$ (left) and for different values of spin parameter with $\alpha=0.1$ (right).}
\label{Luminosity}
\end{figure*}

The effect of PFDM parameter on the spectral energy distribution $\nu L(\nu)$ of thin disk have also displayed in Fig.~\ref{Luminosity}. Similar to the case of radiant energy flux and disk temperature, one may conclude that an accretion disk around a rotating black hole in the presence of PFDM is getting more luminous by increasing $\alpha$ (left panel), while for a fixed value of $\alpha$ the deviation of $L(\nu)$  from that of a static PFDM black hole becomes more pronounced with increasing  $a$ (right panel). Also, we see that with increasing dark matter parameter, the cut-off frequencies, for which the maximum luminosity is obtained, shift to higher values.

Finally, let us define the accretion efficiency $\epsilon$, as the capability of the central object to convert rest mass into outgoing radiation, which can be calculated by knowing the specific energy in the ISCO radius
\begin{equation}
\epsilon=1-\tilde{E}_{\rm isco}.
\end{equation}
In Table~\ref{P1-t1}, we have displayed the numerical results of maximum values of the energy flux, temperature distribution and emission spectra and also radiative efficiency of a rotating black holes in PFDM for different values of spin and dark matter parameters.

\begin{table*}
 \caption{The numerical values of $r_{\rm{isco}}$, the maximum of energy flux, maximum of temperature, the critical frequency, maximum emission spectrum, and efficiency of an accretion disk in PFDM black hole.}
 \label{tab:natbib}
 \begin{tabular}{llllllll}
  \hline
  $a$ & $\alpha$ & $r_{\rm{ISCO}}/M$ &$ F_{\rm{max}}(r) $&$T_{\rm{max}}(r) $&$\nu_{c}$ & $ \nu L_{\rm{max}}(\nu) $&$\epsilon$\\
  \hline
  &0.01&5.8489&$1.337\times{10^{12}}$&$12.391\times10^3$&$7.351\times{10^{14}}$&$1.811\times{10^{37}}$&0.0563\\
  0&0.05&5.4978&$1.539\times{10^{12}}$&$12.835\times10^3$&$7.565\times{10^{14}}$&$1.922\times{10^{37}}$&0.0528\\
  &0.1&5.2256&$1.678\times{10^{12}}$&$13.117\times10^3$&$7.713\times{10^{14}}$&$1.955\times{10^{37}}$&0.0480\\
  \hline
  &0.01&4.4622&$4.274\times{10^{12}}$&$16.571\times10^3$&$8.826\times{10^{14}}$&$3.591\times{10^{37}}$&0.0747\\
0.4&0.05&4.1105&$5.394\times{10^{12}}$&$17.562\times10^3$&$9.231\times{10^{14}}$&$3.933\times{10^{37}}$&0.0720\\
&0.1&3.8388&$6.397\times{10^{12}}$&$18.327\times10^3$&$9.577\times{10^{14}}$&$4.129\times{10^{37}}$&0.0674\\
\hline
  &0.01&2.733&$3.779\times{10^{13}}$&$28.574\times10^3$&$1.225\times{10^{15}}$&$1.059\times{10^{38}}$&0.1256\\
0.8&0.05&2.3262&$6.954\times{10^{13}}$&$33.278\times10^3$&$1.367\times{10^{15}}$&$1.349\times{10^{38}}$&0.1336\\
&0.1&2.0043&$1.224\times{10^{14}}$&$38.336\times10^3$&$1.517\times{10^{15}}$&$1.671\times{10^{38}}$&0.1399\\
  \hline
 \end{tabular}
 \label{P1-t1}
\end{table*}

\section{Image of disks around a rotating black hole in perfect fluid dark matter}
\label{5-image}
In this section, we investigate the image of a thin accretion disk around a rotating black hole in PFDM as seen by a distant observer. To this end, we construct the image of the disk around black hole using numerical ray-tracing technique \citep{Chen:2015cha}. First, we consider a black hole surrounded by an accretion disk and an observer at a distant $D$ from the black hole with inclination angle $i$, (for more details, see the appendix of \citep{Bambi:2016sac}). The image plane for distant observer is described with Cartesian coordinates system $(X,Y,Z)$ where $(X,Y)$ are at the image plane, and the $Z$-axis is perpendicular to the image plane.

We first write the photon initial conditions in the image plane of distant observer. Let us consider the photon at the image plane at the position $(X_0,Y_0,0)$ with 3-momentum $\bmath{k_0}=-k_0\hat{Z}$ perpendicular to the image plane. The initial conditions of photon that reaches the image plane, in the spherical-polar system are given by
\begin{eqnarray}
r^2&=&D^2+X_0^2+Y_0^2,\nonumber\\
\cos \theta_0&=&\frac{D \cos i+ Y_0 \sin i}{r_0},\nonumber\\
\tan \phi_0&=&\frac{X_0}{D\sin i-Y_0\cos i}.
\end{eqnarray}

Also the orthogonality condition uniquely determines the 4-momentum of the photon, which in the spherical coordinates system is
\begin{eqnarray}
k_0^r&=&-\frac{D}{r_0}|\bmath{k_0}|,\nonumber\\
k_0^\theta&=&\frac{\cos i-(Y_0 \sin i+D\cos i)\frac{D}{r_0^2}}{\sqrt{X_0^2+(D\sin i-Y_0\cos i)^2}} |\bmath{k_0}|,\nonumber\\
k_0^\phi&=&\frac{X_0 \sin i}{X_0^2+(D\sin i-Y_0\cos i)^2} |\bmath{k_0}|.
\end{eqnarray}

Using the above initial conditions, we integrate the geodesic equation with backward ray-tracing method in time, from point $(X_0,Y_0,0)$ at the image plane of distant observer to emission points
\begin{equation}
\frac{d^2x^{\mu}}{d\tau^2}+\Gamma^{\mu}_{\nu\rho}\frac{dx^{\nu}}{d\tau}\frac{dx^{\rho}}{d\tau}=0,
\end{equation}
where $\tau$ is the affine parameter. We integrate the geodesic equations using the Runge-Kutta method in MATLAB. Given that the quasar X-ray emission has been observed to follow a power law, we consider the radiation intensity profile emitted from the disk surface as follows
\begin{eqnarray}
I(\nu,\mu,r)\propto\frac{1}{r^n}\frac{\omega(\mu)}{\nu^{\Gamma-1}},
\end{eqnarray}
where $\nu$ is the photon frequency in the rest frame and $\mu$ is the cosine between the photon 4-momentum and the normal vector of the disk measured by the comoving observer,  $\omega(\mu)$ is the angular-dependence of intensity profile which is assumed to be axially symmetrical for simplicity in \citep{Chen:2015cha}, where $\omega(\mu)$  is that mentioned in Chandrasekhar's book \citep{1960ratr.book.....C}. Also, $\Gamma$ and $n$ denote the photon index and the radial steepness of the profile, respectively.

\begin{figure*}
\begin{multicols}{1}
\includegraphics[width=\columnwidth]{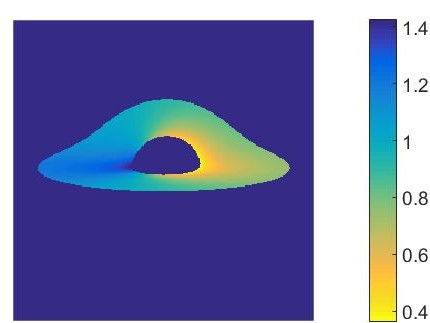}\par
\includegraphics[width=\columnwidth]{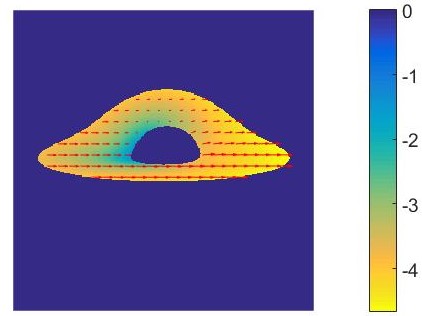}\par
\end{multicols}
\caption{Ray-traced redshifted image (left panel) and intensity and polarization profile (right panel) of a lensed accretion disk around Kerr black hole. The inclination angle is set to $i=80^{\circ}$ and spin parameter is $a=0.5$.}
\label{Kerr1}
\end{figure*}

 \begin{figure*}
\begin{multicols}{1}
\includegraphics[width=\columnwidth]{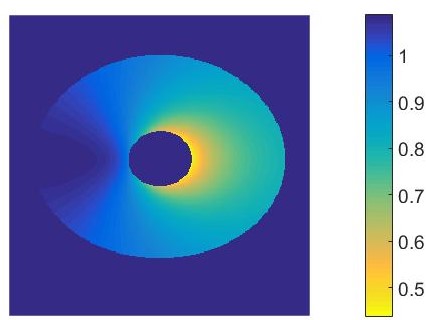}\par
\includegraphics[width=\columnwidth]{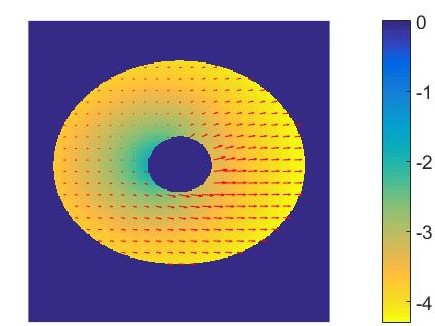}\par
\end{multicols}
\caption{Ray-traced redshifted image (left panel) and intensity and polarization profile (right panel) of a lensed accretion disk around Kerr black hole. The inclination angle is set to $i=40^{\circ}$ and spin parameter is $a=0.5$.}
\label{Kerr2}
\end{figure*}

\begin{figure*}
\begin{multicols}{1}
\includegraphics[width=\columnwidth]{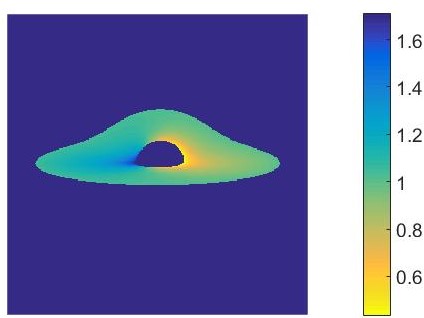}\par
\includegraphics[width=\columnwidth]{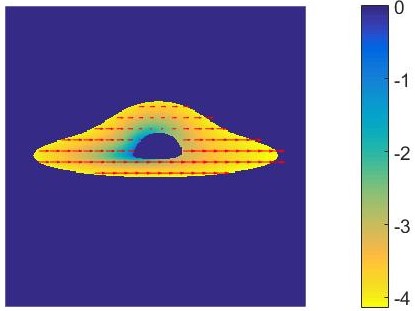}\par
\end{multicols}
\caption{Ray-traced redshifted image (left panel) and intensity and polarization profile (right panel) of a lensed accretion disk around rotating black hole in PFDM. The inclination angle is set to $i=80^{\circ}$, the spin parameter is $a=0.5$ and dark matter parameter is $\alpha=0.2$.}
\label{PFDM1}
\end{figure*}

 \begin{figure*}
\begin{multicols}{1}
\includegraphics[width=\columnwidth]{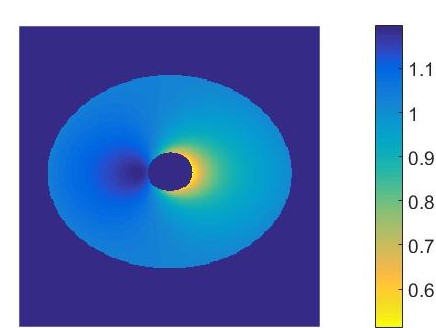}\par
\includegraphics[width=\columnwidth]{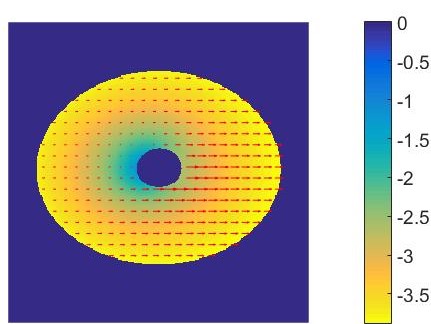}\par
\end{multicols}
\caption{Ray-traced redshifted image (left panel) and intensity and polarization profile (right panel) of a lensed accretion disk around rotating black hole in PFDM. The inclination angle is set to $i=40^{\circ}$, the spin parameter is $a=0.5$ and dark matter parameter is $\alpha=0.2$.}
\label{PFDM2}
\end{figure*}
In Fig.~\ref{Kerr1} and Fig.~\ref{PFDM1}, we present the ray-traced redshifted image of an accretion disk (left panel), and also its intensity and polarization profile  (right panel) around a Kerr black hole with $\alpha=0$, and around a  rotating black hole  with $\alpha=0.2$, spin parameter $a=0.5$  and inclination angle $i=80^{\circ}$.  Fig.~\ref{Kerr2} and Fig.~\ref{PFDM2} are similar to Fig.~\ref{Kerr1} and Fig.~\ref{PFDM1} but plotted at an inclination angle $i=40^{\circ}$. In ray-traced redshifted images, the color bar denotes the degree of the redshift of  radiated light emitted from  the accretion disk. In addition, it also represents the intensity of radiation on the accretion disk for intensity and polarization profiles. Also, the length and direction of arrows seen in intensity and polarization profiles show the size and direction of polarization respectively.

As is well known, some relativistic factors such as light bending, gravitational redshift and Doppler effect change the emission source frequency which leads to  changes of the observed image, as shown in the figures. Such changes are manifest in the hat-like structures (with high inclination angle) and are caused by  gravitational lensing effects of the black hole.

The figures also show Doppler blueshift contributions caused by the rotation of the disk in the left half of the plane which can exceed the overall gravitational redshift due to the presence of black hole, so the corresponding photons have a resulting blueshift (left panels). Also, the intensity distribution is concentrated in a small region near the black hole on the left side of the accretion disk (right panels).

Comparing Fig.~\ref{Kerr1} and Fig.~\ref{Kerr2} for the Kerr space-time without dark matter with Fig.~\ref{PFDM1} and Fig.~\ref{PFDM2} for the rotating black hole in the presence of PFDM shows that the central shadow area of the accretion disk gradually decreases by increasing $\alpha$. Since the space-time under consideration has non-zero angular-momentum, both gravitational Faraday rotation and Thomson scattering affect the photon polarization. Also, with increasing $\alpha$ the size and orientation of polarization in the right panels of the figures decrease so that the Kerr space-time with $\alpha=0$ has the most pronounced change of polarization vector in a small region on the disk near the black hole. Comparing the images for the Kerr black hole at two different inclination angles and also comparing the plots for the Kerr black hole ($\alpha=0$) with rotating black holes surrounded by PFDM ($\alpha\neq0$) shows that the inclination angle has a remarkable effect on both the ray-traced redshifted image  and the intensity and polarization profiles of an accretion disk, while the effect of dark matter parameter $\alpha$ (for $\alpha<\alpha_c$ considered  here) is small.

\section{Conclusions}
\label{6-conclusion}
Accretion disks around supermassive black holes are the primary source of gravitational information in strong gravity regimes and a representative of the geometry of space-time around them. The study of luminosity of accretion disks in the space-time of a static black hole surrounded by dark matter  and  of black holes surrounded by dark matter with isotropic/anisotropic pressure has been  done in \citep{Boshkayev:2020kle,Boshkayev:2022vlv,Kurmanov:2021uqv} respectively. Since astrophysical black holes are expected to be rotating due to the accretion effects, in the present work we have studied thin accretion disk properties around rotating black holes surrounded by PFDM \citep{Xu:2017bpz}.

Considering rotating black holes in PFDM, we studied the properties of thin relativistic accretion disks by employing the Novikov-Thorne model. First, we obtained the physical quantities of interest, namely the effective potential $V_{\rm eff}$, specific angular momentum $\tilde{L}$,  specific energy $\tilde{E}$, angular velocity $\Omega$, and  ISCO radius for test particles moving in circular orbits around the black hole. Since the ISCO equation does not have an analytical solution, we numerically calculated the ISCO radius for both spin, $a$, and dark matter, $\alpha$, parameters. The analysis of the ISCO radius shows that for a given value of $\alpha$  the ISCO radius decreases with increasing spin parameter, while at a fixed spin parameter the ISCO radius decreases up to a critical value of $\alpha$ and then increases with increasing $\alpha$.

To calculate the heat emitted by the accretion disk, we numerically obtained the energy flux $F(r)$, temperature $T(r)$, and luminosity distribution $L(\nu)$ and plotted their profiles. The decrease in ISCO radius shows that in the presence of dark matter the strength of gravitational field decreases with $\alpha$ so that the inner edge of the disk, which is located at the ISCO radius in Novikov-Thorn model, becomes close to the event horizon. Therefore, in comparison with rotating black holes in the absence of PFDM, the energy flux, temperature and luminosity over the disk surface increase which is consistent with the results of Table~\ref{P1-t1}. In summary it can be said that thin accretion disks around rotating black holes surrounded by PFDM are hotter and more luminous than Kerr black holes without dark matter. It is therefore plausible to conclude that all the real mass measured by luminosity is not related to  baryonic mass. It is, however,  possible that some fraction of this luminosity is related to accretion disks around compact objects.

We also investigated accretion disk images around rotating black holes in PFDM using a ray-tracing code written in MATLAB \citep{Chen:2015cha}. We plotted the ray-traced redshifted images, intensity and polarization profiles of  accretion disks for  Kerr black holes in the absence of dark matter and also for rotating black holes surrounded by PFDM. Comparing the corresponding results show that with increasing  $\alpha$ the central shadow area decreases while the polarization vector is susceptible to less rotation. Also, the results show that the effect of $\alpha$ is negligible in comparison with the inclination angle.

\section*{Acknowledgements}
We would like to thank H. R. Sepangi reading  the manuscript and useful comments. Also, we thank the anonymous referee for the useful comments that helped to improve our presentation.

\section*{Data Availability}
No new data were generated or analysed in support of this research.

%%%%%%%%%%%%%%%%%%%%%%%%%%%%%%%%%%%%%%%%%%%%%%%%%%

\bibliographystyle{mnras}
\bibliography{bibl}

\label{lastpage}
\end{document}